\documentstyle[12pt]{article}

\newcommand{\beq}{\begin{eqnarray}}
\newcommand{\eeq}{\end{eqnarray}}

\begin{document}

\begin{flushright}
PSI-PR-94-22\\
UCY-PHY-94/3\\
CERN-TH.7382/94\\
\end{flushright}
\begin{center}

{\bf
\large
Beautiful Baryons  from Lattice QCD \footnote{work
supported in part by DFG grant Schi 257/3-2.} \\}
\end{center}
\normalsize
\medskip

\begin{center}
C. Alexandrou$^{a}$, A. Borrelli$^{b}$, S. G\"usken$^{c}$,  F. Jegerlehner$^b$,
K. Schilling$^{d,c}$, G.~Siegert$^c$, R.~Sommer$^d$\\
\end{center}
$^a$ {\it Department of Natural Sciences, University of Cyprus, Nicosia,
       Cyprus\\}
$^b$ {\it Paul Scherrer Institute, CH-5232 Villigen PSI, Switzerland\\}
$^c$ {\it Physics Department, University of Wuppertal
              D-42097 Wuppertal, Germany\\}
$^d$ {\it CERN, Theory Division, CH-1211 Geneva-23, Switzerland}\\

\centerline{{\bf Abstract}}
\noindent
We perform a
lattice study
of heavy baryons, containing one ($\Lambda_b$) or two $b$-quarks ($\Xi_b$).
Using the quenched approximation we obtain for the
mass of  $\Lambda_b$
$$
 M_{\Lambda_b}= 5.728 \pm 0.144 \pm 0.018 {\rm GeV}.
$$
The mass splitting between the $\Lambda_b$ and the B-meson is found to
increase
by about 20\%
if  the light quark mass is varied from the chiral limit to the strange
quark mass.

\noindent
\newpage

\section{Introduction}
\label{sec:intro}
Considerable progress has been achieved in the computation of low
lying hadronic states from lattice QCD (LQCD). It turned out that {\it
quenched} LQCD  reproduces the hadron spectrum in the
light quark sector surprisingly well, once finite
volume corrections are carefully taken into account~\cite{Weingarten}.
This holds in particular for the ratio $m_N/m_{\rho}$, which
has been a notorious problem for LQCD over quite some years.

Beauty physics is attracting much attention
because the origin of CP violation in the B-system is still an open question.
The investigation of such a system is a great challenge to LQCD.
Considerable progress has been made in the lattice studies of
heavy-light mesons like D- and B-mesons~\cite{revs}.

Heavy-light systems from the {\it baryonic} sector so far have been
studied with lattice methods exploratively in the limit
of infinite heavy quark mass \cite{sta,martin}.

Throughout the current study the mass of the heavy quark
has been kept finite.
We will present results for the mass of
$\Lambda_b$, a baryon composed of a $b$-quark and two light quarks, as
well as for the mass of $\Xi_b$, a baryon composed of two $b$-quarks
and one light quark.

As in the case of heavy-light mesons, the most dangerous source of
systematic error stems from the fact that one is forced --- in order to
enter the region of heavy quark masses --- to actually push the heavy
mass close to the very limit of current lattice resolutions.  We will
attempt, however, to keep control on these effects of a finite
lattice spacing $a$ by a variety of precautions:

1. We avoid to compute the masses of the $\Lambda_b$ and the $\Xi_b$
directly, but rather calculate the mass splittings $\Delta_{\Lambda} =
M_{\Lambda_b} - M_B$ and $\Delta_{\Xi} = M_{\Xi_b} - 2 M_B$,
with respect to the B-meson mass $M_B$. These
splittings do not depend on the heavy quark mass in the infinite mass
limit and are therefore less prone to contamination by finite $a$
effects in the $b$ and $c$ quark mass regions.

2. We monitor the dependence of the splittings on the
lattice spacing for our three $\beta$ values, $\beta =
5.74,6.0,6.26$. This enables  us (a) to check the assumption of weak $a$
dependence and (b) to perform an $a \rightarrow 0 $ extrapolation.

3. We will not calculate directly the mass splittings
too close to the $b$~quark mass. Instead we stop the calculation at
approximately twice the charm quark mass and then extrapolate our
data to the $b$ mass.

Moreover, we investigate finite size effects at $\beta=5.74$, on three
lattices of spatial volumes $8^3, 10^3 $ and $12^3$, in lattice units.
Our  lattice parameters are detailed  in table~1.  To
obtain a good signal for the ground state, we use smeared gauge
invariant interpolating quark fields~\cite{sme},
defined for the standard Wilson action.

\begin{table}[hbtp]
\label{lpar}
\begin{displaymath}
\begin{array}{|c|c|c|c|c|}
\hline
 {\rm N_S} & {\rm N_T} &  {\rm no. \>configs.} &\beta & a\>m_{\rho} \\
\hline
 8 & 24 & 175 & 5.74 & 0.542 \pm 0.014 \\
 10 & 24 & 213 & 5.74 &  \\
 12 & 24 & 113 & 5.74 & \\
\hline
 12 & 36 & 204 & 6.00 & 0.355 \pm 0.016 \\
\hline
 18 & 48 & 67 & 6.26 & 0.260 \pm 0.014 \\
\hline
\end{array}
\end{displaymath}
\caption{Lattice parameters: space and time extents $N_S$ and $N_T$,
number of configurations, $\beta$ and the  $\rho$-mass  in lattice units.}
\end{table}

\section{Smearing Techniques and Volume Effects}
\label{sec:smvol}

For the reasons given above, we will compute the baryonic masses with
reference to the mass of the B-meson, by proper combinations that
would eliminate the $b$-quark mass in the
heavy quark limit. So we consider the splittings
$\Delta_{\Lambda} = M_{\Lambda_b}- M_B$ and
$\Delta_{\Xi}=M_{\Xi_b}-2 M_B$, respectively, in the single and double
beauty sectors.
In the first case, we need to compute the correlators, for the
heavy-light pseudoscalar meson,
\beq
C_P(t) = \sum_{\vec{x}} \left\langle \left( \bar{Q}(x) \gamma^5 q^{J}(x)
\right)
         \left( \bar{q}^{J}(0)  \gamma^5 Q(0) \right) \right\rangle,
\eeq
and for the $\Lambda$ baryon  \cite{Boch}
\beq
 C_{\Lambda}(t) = \sum_{\vec{x}} \left\langle
    \left(  \epsilon^{abc} {Q}_a(x)
          \left( q_b^{J}(x) {\rm C} \gamma_5 q^J_c(x) \right)
   \right) \right. \nonumber \\ \left.
    \left(  \epsilon^{abc} {Q}_a(0)
          \left( q_b^{J}(0) {\rm C} \gamma_5 q^J_c(0) \right)
   \right)^{\dagger} \right\rangle,
\eeq
\noindent
where $Q(x)=Q({\vec{x}}, t)$
is the heavy quark field, and $q^{J}(x)=q^{J}({\vec{x}}, t)$ is the
light one, to which smearing of type $J$~\cite{sme} has been
applied\footnote{The correlator for the $\Xi$ is obtained from the one
for the $\Lambda$, by replacing the light quark $q_b^{J}$ by a heavy
one $Q_b$.}. $C$ is the charge conjugation operator given by $C = i \gamma^4
\gamma^2$.

Given the lattice results for these correlators, we
perform  a direct fit to their ratio
\beq
 R_{\Lambda}(t)=\frac{C_{\Lambda}(t)}{C_P(t)}\rightarrow
                A e^{-\Delta_{\Lambda}t}
\eeq
in the large $t$ limit.

It is by now well known that  smearing \cite{sta,Fer,UKQCD}
is crucial  to obtain a decent overlap of the operators with the
ground state. In this work, we make use of the experience acquired
previously while  computing properties of the  heavy-light
pseudoscalar states~\cite{sta}, where we found gauge-invariant
`Gaussian type' wave functions (of r.m.s radius $0.3$~fm)
to provide sufficient overlap. The smearing was applied to the
light quark source in the mesonic case. In this
 study, we are
using precisely this procedure for the heavy-light baryons as well,
without any further optimization.
\par
In order to establish ground state dominance we look for a plateau
in the local mass of the ratio $R_{\Lambda}$.
In fig.~1a we show as an example the local mass of $R_{\Lambda}$,
for the two largest lattices:
the solid line shows the
fitted value for the plateau.
Fig.~1b shows the corresponding local mass for
the ratio $R_{\Xi}(t)=C_{\Xi}(t)/C_P^2(t)$
used to determine $\Delta_{\Xi}$.
These figures show the quality of the plateaus
 for representative $\kappa$ values. Worse quality is found only in a few
 cases, and it resulted in larger statistical errors;
 for instance the two largest $\kappa$ values at $\beta=5.74$
 given in table~3.

The pseudoscalar mass was extensively studied in ref. \cite{fb} and the
values have been taken
from there.

For checking the  finite volume effects, we  computed
 $\Delta_{\Lambda}$ on three lattices with ${\rm N_T}=24$,
$\beta=5.74$ and sizes ${\rm N_S}=$8, 10, 12, for $\kappa_l=0.156$,
and $\kappa_h=$ 0.125, 0.140, 0.150.
 We compare the results for the splitting
 in table~\ref{vol}. The values exhibit deviations of at most
$4 \%$ .
In fig.~2 we plot the dimensionless ratio
${\Delta_{\Lambda}/{m_{\rho}}}$ as a function of $L$, the lattice size
in units of $m_{\rho}$; as can be seen, for these values of
$\kappa_h$, the finite size effects of this ratio are smaller than our
statistical errors.

In the following, we will fix the volume to about 1~fm
(which corresponds to $N_S \simeq $8, 12 and 18 for $\beta=5.74$, 6.0 and
6.26 respectively) and carry out a detailed study of the extrapolation
to the continuum limit. As a tribute to possible finite size effects
we will add the above maximal variation  of $4\%$ as an
uncertainty to all results.

\begin{table}[hbtp]
\label{vol}
\begin{displaymath}
\begin{array}{|c|c|c|c|c|}
\hline
 \kappa_l & \kappa_h & \Delta_{\Lambda}, 8^3\times 24 &
                       \Delta_{\Lambda}, 10^3 \times 24 &
                       \Delta_{\Lambda}, 12^3 \times 24 \\
\hline
 0.156 & 0.125 & 0.564 \pm 0.008 & 0.554 \pm 0.005 & 0.569 \pm 0.005 \\
\hline
   & 0.140 &0.592 \pm 0.007 & 0.575 \pm 0.005 & 0.588 \pm 0.004 \\
\hline
   & 0.150 & 0.621 \pm 0.007 & 0.601 \pm 0.005 & 0.612 \pm 0.006 \\
\hline
\end{array}
\end{displaymath}
\caption{Results for three different spatial lattice extents,
${\rm N_S}=$8, 10, and 12, ${\rm N_T}=24$, $\beta=5.74$.}
\end{table}
\begin{table}[p]
\label{res1}
\begin{displaymath}
\begin{array}{|c|c|c|c|c|c|}
\multicolumn {6}{c}{8^3 \times 24, \quad \beta=5.74} \\
\hline
   \kappa_h & a\Delta_{\Lambda} & a\Delta_{\Lambda}(m_s) &
 a\Delta_{\Xi} & a\Delta_{\Xi}(m_s) &
 \Delta_{\Lambda}(m_s) / \Delta_{\Lambda}(m_u)\\
\hline
  0.06 & 0.25 \pm 0.13 & 0.37 \pm 0.08 & -0.24 \pm 0.02
& -0.31 \pm 0.02
 & 1.46 \pm 0.40\\
\hline
 0.09 & 0.34 \pm 0.10& 0.42 \pm 0.06&
 -0.17 \pm 0.02 & -0.23 \pm 0.02
& 1.23 \pm 0.18 \\
\hline
 0.125   & 0.38 \pm 0.04 & 0.46 \pm 0.02 & -0.12 \pm 0.02 & -0.17 \pm 0.01
& 1.20 \pm 0.07 \\
\hline
 0.14   & 0.42 \pm 0.02 & 0.50 \pm 0.01 & -0.01 \pm 0.01 & -0.009 \pm 0.009
& 1.18 \pm 0.04 \\
\hline
 0.15 & 0.47 \pm 0.04 & 0.54 \pm 0.02 & 0.09 \pm 0.01 & -0.080 \pm 0.008
& 1.13 \pm 0.05 \\
\hline
\multicolumn {6}{c}{} \\
\multicolumn {6}{c}{12^3 \times 36, \quad \beta=6.00} \\
\hline
   \kappa_h & a\Delta_{\Lambda} & a\Delta_{\Lambda}(m_s) &
 a\Delta_{\Xi} & a\Delta_{\Xi}(m_s) &
 \Delta_{\Lambda}(m_s) / \Delta_{\Lambda}(m_u)\\
\hline
 0.1   & 0.22 \pm 0.02 & 0.28 \pm 0.01 & -0.14 \pm 0.02 & -0.18 \pm 0.01
& 1.23 \pm 0.05 \\
\hline
 0.115  & 0.21 \pm 0.02 & 0.28 \pm 0.01 & -0.10 \pm 0.01 & -0.147 \pm 0.008
&1.29 \pm 0.08\\
\hline
 0.125 & 0.25 \pm 0.01 & 0.300 \pm 0.008 & -0.09 \pm 0.01 & -0.130 \pm 0.007
& 1.19 \pm 0.03\\
\hline
 0.135 & 0.27 \pm 0.01& 0.310 \pm 0.007
 & -0.06 \pm 0.01 & -0.101 \pm 0.006
& 1.17 \pm 0.03 \\
\hline
 0.145 & 0.31 \pm 0.02 &  0.348 \pm 0.007 & 0.01 \pm 0.01 & -0.049 \pm 0.009
 & 1.11 \pm 0.04 \\
\hline
\multicolumn {6}{c}{} \\
\multicolumn {6}{c}{18^3 \times 48, \quad \beta=6.26} \\
\hline
   \kappa_h & a\Delta_{\Lambda} & a\Delta_{\Lambda}(m_s) &
 a\Delta_{\Xi} & a\Delta_{\Xi}(m_s) &
 \Delta_{\Lambda}(m_s) / \Delta_{\Lambda}(m_u) \\
\hline
 0.09   & 0.15 \pm 0.01 & 0.188 \pm 0.009 & -0.07 \pm 0.01 & -0.107 \pm 0.008
& 1.25 \pm 0.07\\
\hline
 0.10 & 0.15 \pm 0.02 & 0.19 \pm 0.01 & -0.073 \pm 0.008 & -0.110 \pm 0.006
& 1.29 \pm 0.07 \\
\hline
 0.12 & 0.16 \pm 0.01 & 0.201 \pm 0.007 & -0.062 \pm 0.06 & -0.010 \pm 0.004
& 1.25 \pm 0.04 \\
\hline
 0.135 & 0.18 \pm 0.01 & 0.216 \pm 0.008 & -0.032 \pm 0.005 & -0.071 \pm 0.004
& 1.21 \pm 0.04 \\
\hline
 0.145   & 0.21 \pm 0.01 & 0.243 \pm 0.009 & 0.022 \pm 0.006 & -0.024 \pm 0.005
& 1.14 \pm 0.04 \\
\hline
\end{array}
\end{displaymath}
\caption{The mass splittings $\Delta_{\Lambda}$ and $\Delta_{\Xi}$
at the chiral limit and at the strange quark mass in lattice units for
$\beta=5.74, 6.00, 6.26$. The
last column gives the ratio
$\Delta_{\Lambda}(m_s) / \Delta_{\Lambda}(m_u)$ for the same $\beta$
values.}
\end{table}

\section{Continuum Limit}
\label{sec:cont}
{\bf $\Lambda$ Splitting.}
 The results for $\Delta_{\Lambda}$ at the three $\beta$ values
 5.74, 6.00 and 6.26 at fixed volume
  are listed in table~3, all in lattice units.
 The extrapolations to  $u$ and $s$-type light quarks, ---  for given
heavy quark  $\kappa_h$ ---
are performed as described in \cite{sta}.

Since we are evaluating masses, it is natural to use $m_{\rho}$ to set
the lattice scale.
  The values of $m_{\rho}$ for the various lattices have
been listed in table~1; the experimental value used is: $m_{\rho}=768$~MeV.

In fig.~3a we plot $\Delta_{\Lambda}$ in GeV extrapolated to the chiral
limit as
a function of ${1/{M_P}}$ in ${\rm GeV}^{-1}$.
Within the statistical precision achieved in this computation,
the points show no dependence  on $a$, although $a$ is varied by about
a factor two (cf. Table 1).

In the continuum, the $1/M_P$ expansion for $\Delta_{\Lambda}$  gives
\beq  \Delta_{\Lambda}^{\rm cont.}(M_P) = c_0+{c_1\over{M_P}} +
O({M_P}^{-2}) \; .
\label{an1}
\eeq

Assuming no dependence on the lattice spacing, this form can be fitted
directly
to the points in fig. 3a yielding  $\Delta_{\Lambda_b}= 431(28)~{\rm MeV} $ at
the
mass of the B-meson. This result is included in the figure as the inverted
triangle. The error bar of this point does not account for the fact that the
simulation results exclude an $a$-dependence only within their precision.

A realistic error that includes the uncertainty of extrapolating the
lattice data  to the continuum is obtained by allowing for the
leading  $a$-dependence  at each value of $M_P$~\cite{fb}:
 We start out from a selected  value of $M_P$ (in physical units)
and interpolate the lattice results from  each (fixed) $\beta$-value
to the value of $M_P$.
This enables us
to compare  $\Delta_\Lambda$ at different values of $a$.
A subsequent  linear extrapolation in $a$ will then
yield  the continuum estimate for
 $\Delta_\Lambda$,  at the chosen  physical value of $M_P$.

Reiteration of the procedure on a set of  masses $M_P$
determines  the numerical dependence of
$\Delta_{\Lambda}^{\rm cont.}$ on $M_P$. In order to
remain in compliance with the
assumption of linearity in $a$,
we used only a subset of the data listed in
table~3, by excluding the data for the two heaviest masses at any value of
$\beta$.

A final fit of the continuum limit values $\Delta_{\Lambda}^{\rm cont.}$
to eq.(\ref{an1}) yields
\beq
\Delta_{\Lambda_b}= 458 \pm 144 \pm 18 \> {\rm MeV} \;.
\label{continuous_mass}
\eeq
Here the first error is purely statistical and stems from a full jacknife
analysis of our data. The second error represents the $4\%$ uncertainty
due to finite volume effects.
Together with
   the known experimental value $M_B=5.27$~ GeV,
 we obtain an estimate for the mass of
$\Lambda_b$~:
\beq
 M_{\Lambda_b}= 5.728 \pm 0.144 \pm 0.018 \> {\rm GeV}.
\label{mass_lambda}
\eeq
We note in passing, that the parameter $c_0$ is in agreement
with the estimates obtained directly in the static approximation~\cite{sta},
albeit within the larger uncertainties of the latter.

\par In the same way we obtain the mass splitting $\Delta_{\Lambda_b}(m_s)$
where the light quark mass is extrapolated to the strange quark mass. The
data fitted are shown in fig.~3b, and the result of the extrapolations
described above is
\beq
\Delta_{\Lambda_b}(m_s) = 597 \pm 91 \pm 24 \> {\rm MeV}~,
\label{dels_full}
\eeq
with the same meaning of errors.
Instead of looking directly
 at  the splitting at the strange quark mass,
one may consider the
ratio
\beq
  r=\frac{\Delta_{\Lambda}(m_s)}{\Delta_{\Lambda}(m_u)},
\label{ratio}
\eeq
which is
expected to have an even weaker $a$-dependence.
Indeed the ratio for the different
$\beta$ values shown in
fig.~3c follows a ``universal curve" with quite small statistical errors.
A linear fit to the (weak)
mass dependence yields
\beq
 r=1.25  \pm 0.03  \pm 0.05.
\eeq
This ratio represents the change in the $\Lambda_b$ mass when replacing
the $u$ quark by an $s$ quark.


{\bf $\Xi$ Splitting.}
The mass splitting for the $\Xi$ baryon is investigated using the same
techniques
as for the $\Lambda$.
Table~3 displays the results after extrapolation of the light quark
to the $d$ and $s$ mass  respectively. We
then convert
all data to physical units, and plot $\Delta_{\Xi}$ vs. $1/M_{P}$ at
the chiral limit in fig.~4. In this case the data show a
statistically  significant
$a$-dependence, especially for heavy masses.
We extrapolate to the continuum limit
as above for  pseudoscalar masses $M_P$ below
the charm mass. We use the
same ansatz as in the case of $\Delta_{\Lambda}$ to fit the  $1/M_P$
dependence of $\Delta_{\Xi}$ and  obtain:
\beq
         \Delta_{\Xi_b}= -90 \pm 170 {\rm \ MeV \ },
\eeq
This mass shift
together with the experimental value for $M_B$ determines
\beq
                M_{\Xi_b} = 10.45 \pm 0.17 {\rm  \ GeV \ }~~,
\eeq
for the physical $\Xi_b$--mass.
\par Another possibility to compute the $\Xi_b$ mass, is to consider the
ratio $R~'_{\Xi}(t)=C_{\Xi}(t)/C_P^{{\bar h}h}(t)$, which should yield the
splitting between the $\Xi_b$ mass and the $\Upsilon$-meson mass.
We evaluated this quantity but the resulting
 plateaus turned out to be of  worse quality than those for the
 other channel studied; this is due to the fact that smearing was
 applied only to light quarks, and so the heavy-heavy channels
 suffer from relevant contamination by higher-mass states.

\section{Discussion}

The mass splitting technique is a viable method to
compute the $\Lambda_b$ mass on the lattice. Both the lattice spacing
dependence of the mass splitting and its dependence on
the heavy quark mass are weak. Thus an extrapolation to the
continuum and to the b--quark mass is possible.
Our actual value
of $ M_{\Lambda_b}= 5728 \pm 144 \pm 18~ {\rm MeV}$ can be compared
to the value 5630 MeV suggested by Martin et al. within  the
naive quark model approach~\cite{martin}.

Experimentally, the determination of the $\Lambda_b$ mass has a
somewhat controversial history ever since 1981~\cite{rosenfeld}.
Recently, the UA1 collaboration has measured the mass from 16 events
in the decay channel $J/\Psi \Lambda$\footnote{It is an open question,
 why this decay channel has not been observed in the CDF
experiment.}.  Their value is~\cite{UA1} $M_{\Lambda_b} = 5640 \pm 50
\pm 30$MeV.  The DELPHI collaboration is presently  quoting~\cite{Delphi} a
preliminary mass value, $M_{\Lambda_b} = 5635^{+38}_{-29} \pm 4$MeV, which
is based on one candidate event in the  $\Lambda _c^+ \pi ^-$
and in the $D^0 p\pi -$ decay modes.
These numbers are in agreement with our result.

 For
$M_{\Lambda_c}$ we obtain $2433 \pm 88 \pm 18$ MeV, which is in rough
accord
with the experimental
value of 2285 MeV~\cite{rosenfeld}.
Looking at the light quark dependence we found  a 20\%
increase  as we lift  the  quark mass from the chiral
limit to the strange quark mass.
Finally, an estimate for the $\Xi_b$ mass is given in eq.(11).

\par Our errors do include -- as the dominant part -- the uncertainty
induced by an extrapolation to the continuum limit. Nevertheless,
it is desirable to further check these extrapolations
through simulations with higher lattice resolutions and/or different
lattice actions.

\section*{Acknowledgements}
We thank the Personal at the Computer Centers CSCS in Manno Switzerland and
J\"ulich, Germany for their support.
KS thanks A.\ Martin for a valuable   discussion.


\section*{Figure captions}
\begin{enumerate}
\item[] {\bf Figure 1a}\\
The $\Delta_{\Lambda}$ local masses, given by $\mu_{\Lambda}^{\rm loc}(t)=
{\rm log}[\frac{R_{\Lambda}(t-a)}{R_{\Lambda}(t)}]$
for $\kappa_l=0.1525$, $\kappa_h=0.125$ at $\beta=6.0$,
for a $12^3 \times 36$  lattice, and for $\kappa_l=0.1492$, $\kappa_h=0.12$
at $\beta=6.26$, for
a $18^3 \times 48$  lattice.

\item[] {\bf Figure 1b}\\
The same as figure 1a, but for $\mu_{\Xi}^{\rm loc}(t)=
{\rm log}[\frac{R_{\Xi}(t-a)}{R_{\Xi}(t)}]$. \\

\item[] {\bf Figure 2}\\
$\Delta_{\Lambda}/m_{\rho}$ is shown vs $L$ for three heavy quark
masses at $\beta=5.74$. The light quark mass was fixed to about twice the
strange quark mass.

\item[] {\bf Figure 3a}\\
The $\Lambda$ mass splitting is shown vs $1/M_P$   in GeV$^{-1}$ at the chiral
limit.
The solid line is a global fit assuming no $a$ effects.
It gives
 the value shown with the inverted triangle at the B-meson mass.
The value obtained after extrapolation to the continuum limit
and to the
B-meson mass, is shown by
the open circle. This extrapolation is discussed in the text.\\

{\bf Figure 3b}\\
As for figure~3a but at the strange quark mass. \\

{\bf Figure 3c}\\
The ratio $\Delta_{\Lambda}(m_s)/\Delta_{\Lambda}(m_u)$ is shown
vs $1/M_P$. The
notation is the same as for figure~3a.

\item[] {\bf Figure 4}\\
The $\Xi$ mass splitting, $\Delta_{\Xi}=M_{\Xi}-2M_{P}$
 is shown vs $1/M_P$   in GeV$^{-1}$ at the chiral limit.
The
notation is the same as for figure~3a.\\

\end{enumerate}


\begin{thebibliography}{99}
\bibitem{Weingarten}{F. Butler, H. Chen, J. Sexton, A. Vaccarino and D.
Weingarten, IBM-Report 1994 (HEP-LAT-9405003)}
\bibitem{revs} {
see e.g.
C. Bernard in {\it  Lattice 93} Nucl. Phys. B (Proc. Suppl.) 34 (1994) 47;
R. Sommer,{\it Beauty Physics in Lattice Gauge
Theory}, preprint DESY 94-011, to be published in Physics Reports C.}
\bibitem{sta}{
 C. Alexandrou, S. G\"usken, F. Jegerlehner, K. Schilling,
 and R. Sommer, Nucl. Phys. B414 (1994) 815.}
\bibitem{martin}
{A.\ Martin and J.M. Richard, Phys.Lett. B185 (1987) 426.}
\bibitem{sme}{S.G\"usken, in {\em Lattice 89}, Nucl. Phys.
 B (Proc. Suppl.) 17 (1990) 361}
\bibitem{Boch}{M. Bochicchio, G. Martinelli, C. R. Allton, C. T.
Sachrajda and D. B. Carpenter, Nucl. Phys. B372 (1992) 403}
\bibitem{Fer}{A. Duncan, E. Eichten and H. Thacker, Nucl. Phys. B (proc. Supp.)
 30 (1993) 441.}
\bibitem{UKQCD}{UKQCD Collaboration, S. Collins et al.,  Nucl. Phys. B
(Proc.~Supp.) 30 (1993) 393.}

\bibitem{fb}{
 C. Alexandrou, S. G\"usken, F. Jegerlehner, K. Schilling, G. Siegert
 and R. Sommer, DESY preprint 93-179, Z. Phys. C in press.}
\bibitem{Aoki}{S. Aoki et al., Nucl. Phys. B (Proc. Suppl.) 34 (1994) 363.}
\bibitem{fb0}{
 C. Alexandrou, S. G\"usken, F. Jegerlehner, K. Schilling,
 and R. Sommer, Nucl. Phys. B374 (1992) 263.}
\bibitem{rosenfeld}
{Particle Data Group, K. Hikasa et al.,  Phys. Rev. D45 (1992) 4365.}
\bibitem{UA1}{C. Albajar et al., UA1 collaboration,
Phys. Lett. B273 (1991) 540.}
\bibitem{Delphi}{M. Battaglia et al., DELPHI collaboration, DELPHI note
DELPHI 94-30 PHYS 363, March 1994.}
\par
\end{thebibliography}
\end{document}